Effects of dopant type and concentration on the femtosecond laser ablation threshold and incubation behaviour of silicon

Reece N. Oosterbeek[1,2,3], Carsten Corazza[1,2,3], Simon Ashforth[1,3,4], M. Cather Simpson[1,2,3,4,*]

[1] The Photon Factory, The University of Auckland, Private Bag 92019, Auckland 1142, New Zealand

[2] School of Chemical Sciences, The University of Auckland, Private Bag 92019, Auckland 1142, New Zealand

[3] The Dodd Walls Centre for Quantum and Photonic Technologies, and The MacDiarmid Institute for Advanced Materials and Nanotechnology, New Zealand

[4] Department of Physics, The University of Auckland, Private Bag 92019, Auckland 1142, New Zealand

* Corresponding author email: c.simpson@auckland.ac.nz

Abstract (150-250 words)

In laser micromachining, the ablation threshold (minimum fluence required to cause ablation) is a key performance parameter and overall indicator of the efficiency of material removal. For pulsed-laser micromachining, this important observable depends upon material properties, pulse properties and the number of pulses applied in a complex manner that is not yet well understood. The incubation effect is one example. It manifests as a change in the ablation threshold as a function of number of laser pulses applied, and is driven by photoinduced defect accumulation in the material. Here we study femtosecond (800 nm, 110 fs, 0.1 to 1 mJ/pulse) micromachining of a material with well-defined initial defect concentrations: doped Si across a range of dopant types and concentrations. The single pulse ablation threshold ($F_{th,1}$) was observed to decrease with increasing dopant concentration, from a maximum of 0.70 J/cm$^2$ (± 0.02) for undoped Si, to 0.51 J/cm$^2$ (± 0.01) for highly N-type doped Si. The effect was greater for N-type doped Si than for P-type, consistent with the higher carrier mobility of electrons compared to holes. In contrast, the infinite pulse ablation threshold ($F_{th,\infty}$) was the same for all doping levels and types. We attribute this asymptotic behaviour to a maximum defect concentration that is independent of the initial defect concentration and type. These results lend insight into the mechanism of multi-pulse, femtosecond laser ablation. (225 words)

Keywords: ablation threshold, doped silicon, ultrafast laser, laser micromachining

1. Introduction

In recent years, significant advances have been made in laser micromachining technologies that have led to the exploration of femtosecond laser pulses in the micromachining industry. Femtosecond laser pulses have the ability to ablate with low pulse energies, because their ultrashort duration gives them very high peak powers – GW or higher. This high peak irradiance means that materials undergo ablation independent of linear optical absorption properties, through multiphoton and other nonlinear interaction mechanisms. Further, because the laser pulses are so short, the transfer of energy from the excited electron plasma to the nuclear lattice of the material is ineffective. Consequently, thermal and other types of collateral damage are greatly reduced when femtosecond laser pulses are used for micromachining [1-3]. Despite these obvious benefits, when compared to nanosecond or continuous wave laser ablation, ultrashort lasers have yet to become widely used in industry, as processing speeds are still too slow [4, 5].

We and others are actively studying ways to improve femtosecond laser micromachining speeds. To evaluate femtosecond laser machining efficiency, we measure the ablation threshold. The ablation threshold is the minimum pulse fluence required to remove material [6-8]. To first order, the laser ablation threshold should be a property of the material. It reflects the ability of the material to hold onto its electrons when irradiated by light, and is thus related to the material's work function. Unfortunately, in practice the phenomenon is not so simple, particularly when lasers with pulse durations of less than about a picosecond are employed.

In these cases, the absorption process becomes highly nonlinear. The pulse interacts with the electron plasma as it being generated, and temporally overlapping ionization cascades eventually lead to the ejection of material. The cross-section for nonlinear absorption depends sensitively upon the pulse characteristics such as duration, energy, frequency and chirp [9]. The detailed structure of the material, including level and type of defects, plays an important role as well [2, 10-14]. The experimentally measured ablation threshold reflects all of these interactions in a complex manner (Figure 1).

Femtosecond laser ablation is known to be heavily dependent on the material's electrical properties, as evidenced by multiple studies showing that the femtosecond laser ablation process occurs via electronic excitation and ionisation [2, 11, 13]. Less heavily studied in recent years have been the effects of incubation. This phenomenon has been described in terms of accumulation of Frenkel-type defects [10] which affect the distribution of charge within the lattice, raising questions about what effect some initial concentration of defects (i.e. dopant atoms) would have on the incubation behaviour.

This study is part of a larger project to understand the complex interactions between the femtosecond pulses and dielectric materials, and to exploit that knowledge to improve laser micromachining efficiency. Here, we maintain the laser pulse characteristics, and examine the laser ablation threshold for a series of well-characterized doped silicon samples. The study focuses in particular upon the impact of defects upon the ablation threshold, by examining incubation effects. For many materials, the laser ablation threshold decreases as larger numbers of successive pulses are applied. This means that the single pulse ablation threshold is usually higher than the ablation threshold measured when multiple pulses are applied to the same location. This phenomenon occurs even if the time interval between pulses is sufficient for thermal equilibration to have occurred. The effect is attributed to defect formation: laser pulses remove material and induce defects in the material that is not ablated; consequently the following pulses are able to ablate the material more easily [10, 12, 15].

Doped silicon provides an ideal material for examination of the effects of initial defects on laser ablation, as it is readily available in a range of doping types and concentrations that are well controlled. Doping of semiconductors such as silicon is commonly carried out for device manufacturing, in order to alter the electrical properties of intrinsic silicon, where both N-type doping (usually with phosphorus) and P-type doping (usually with boron) are used to create charge carriers (electrons and electron holes respectively). The concentration of these dopants has a major impact on the material's electrical properties, decreasing resistivity and narrowing the bandgap [16-23]. These variable bandgaps are well documented and can be calculated with knowledge of the doping type and concentration, and are heavily used in the manufacture of semiconductor devices.

Fang and Cao et al. have previously studied the effects of N-type doping on the femtosecond laser ablation of silicon [24, 25], with results suggesting that doping can reduce the ablation threshold by up to 20%. These results however, only consider N-type doping, across a limited range of pulse numbers. Our hypothesis is that the different carrier mobilities within the lattice will lead to different ablation and incubation behaviour. Here we investigate the effect of both N-type and P-type doping of Si across a large range of pulse numbers (~1–30,000), in order to determine how these parameters affect the laser ablation and incubation behaviour.

2. Materials and Methods

2.1 Laser Setup

Laser ablation experiments were carried out using a femtosecond laser micromachining system - a regeneratively amplified Ti:Sapphire femtosecond laser consisting of an oscillator (Mantis, Coherent Inc., USA) and an amplifier (Legend Elite, Coherent Inc., USA). The 800 nm, 110 fs pulsetrain, with a repetition rate of 1 kHz, was directed through a pulse picker (Series 5046ER, Fastpulse Inc., USA) to control the repetition rate, and an attenuator (Watt Pilot, UAB Altechna, Lithuania) to control the pulse energy. The beam then passed through a mechanical shutter (SH05, Thorlabs Inc., USA) and f=75 mm focusing lens (LA1608-B-ML, Thorlabs Inc., USA). The beam was focused onto the sample, mounted on a 3-axis micromachining stage, consisting of linear translation stages (Thorlabs Inc., USA). This setup is described in Figure 2; all experiments were carried out under atmospheric conditions (1 atm air, room temperature 20-25 ºC).

2.2 Ablation Threshold Measurement

Ablation thresholds were measured using the diagonal scan (D-Scan) method developed by Samad et. al. [26, 27]. In this technique, the sample is first placed in the path of the focused Gaussian laser beam, above the focal point. The sample is then translated along the *y* and *z* axes simultaneously (directions perpendicular and parallel to the optical axis - Figure 3). This process results in the formation of a damage feature with a characteristic "two lobe" shape. By measuring of the maximum damage radius of this feature ($\rho_{max}$, in cm), the ablation threshold can be calculated directly using equation 1 [26], where $F_{th}$ is the ablation threshold (J/cm$^2$), $E_0$ is the pulse energy (J), and *e* and $\pi$ are mathematical constants.

$$F_{th} = \frac{E_0}{e\pi\rho_{max}^2} \qquad (1)$$

To study incubation effects, the number of pulses applied must also be quantified, using equation 2 [27], where *N* is the pulse superposition (i.e. number of pulses applied), *f* is the laser repetition rate (Hz), and $v_y$ is the sample translational speed in the transverse direction (cm/s).

$$N = \frac{\sqrt{\pi} f \rho_{max}}{v_y} \qquad (2)$$

Diagonal scans were performed at translational speeds ranging from 10 – 750 cm/s, and using laser repetition rates of 4 Hz, 10 Hz and 1 kHz, in order access a range of pulse superposition values from ~1-30,000. No difference in the laser-material interaction is expected for the different repetition rates used, because even at the maximum repetition rate used (1 kHz), the pulse separation (1 ms) is sufficient time for any plasma or heat to dissipate – the only lasting effect expected between pulses is permanent or quasi-permanent structural changes. This is confirmed by multiple previous experiments [27-30]. Laser pulse energies from 0.1 to 1 mJ/pulse were used in order to obtain the required D-Scan feature shape in accordance with previous results and published methodology [26, 27].

Ablation features were imaged using an optical profiler (Contour GT-K, Bruker Inc., USA), operated in vertical scanning interferometry mode (VXI), to obtain a surface height plot. Measurements of the maximum damage radius were then performed on these surface plots, using the Vision64 software (Bruker Inc., USA).

2.3 Materials

The samples used for this experiment were polished monocrystalline silicon, <100> orientation, 275 µm thick, supplied by University Wafer Inc., USA. N-type silicon was doped with phosphorus, while P-type silicon was doped with boron. Resistivities were specified to fall within a certain range by the manufacturer, these are summarised in Table 1, along with the corresponding calculated dopant concentration [19-23].

3. Results and Discussion

3.1 Ablation Features

Figure 4 shows examples of the diagonal scan ablation features machined into the silicon surface, measured by optical profilometry. The characteristic diagonal scan profile can be seen, with the maximum damage radius indicated. We observe a clear asymmetry about the centre of the feature. We attribute this to distortion of the beam at the focal point, where a plasma is generated, causing distortion and scattering of the beam after the focal point. This phenomenon is consistent with that seen previously [31, 32]. Because of this phenomenon, we use the maximum damage radius measured from the left lobe only, where the sample is located above the focal point (denoted as *-z* in Figure 4).

3.2 Ablation Thresholds

Incubation data and fitted curves showing the change in ablation threshold with pulse superposition are shown in Figure 5 for each doping level. We observe that incubation is occurring to a high degree in all cases, with ablation thresholds decreasing to 10-20% of the single pulse ablation threshold ($F_{th,1}$) after approximately 1000 pulses ($F_{th,\infty}$). This incubation behaviour has been fitted to the model proposed by Sun et. al. [12], using the MATLAB script made available by the authors [33]. This model describes incubation in terms of two physical mechanisms – the pulse induced change in absorption, and the pulse induced change in critical energy deposition required for ablation. This model is seen in equation 3, where $F_{th,N}$ is the ablation threshold for *N* pulses, $F_{th,1}$ is the single pulse ablation threshold, $F_{th,\infty}$ is the infinite pulse ablation threshold, *N* is the number of pulses applied, $\Delta\alpha/\alpha_0$ is the change in absorption, and $\beta$ and $\gamma$ represent the rate of the change for absorption and critical fluence respectively.

$$F_{th,N} = \frac{F_{th,1} - \left[F_{th,1} - F_{th,\infty}\left(1 + \frac{\Delta\alpha}{\alpha_0}\right)\right]\left[1 - e^{-\gamma F_{(r)}(N-1)}\right]}{1 + \frac{\Delta\alpha}{\alpha_0}\left[1 - e^{-\beta F_{(r)}(N-1)}\right]} \quad (3)$$

Fit curve parameters for each data set are shown in Table 2 along with corresponding $R^2$ values – it appears that this model fits our data well. The ablation threshold of doped silicon has not been comprehensively studied, however our results for undoped silicon are supported by those of Bonse et al. (0.1 – 0.2 J/cm$^2$ after 100 pulses, 800 nm, 100 fs) and Shaheen et al. (0.54 J/cm$^2$ after 20 pulses, 785 nm, 130 fs) [14, 34].

To investigate the ablation and incubation behaviour in more detail, we examined both the single and infinite pulse ablation threshold values obtained for each different doping level and type (Figure 6). A clear trend can be seen for the single pulse ablation threshold ($F_{th,1}$ – Figure 6A), with the undoped silicon having the highest threshold, and a monotonic decrease in ablation threshold as resistivity decreases (dopant concentration increases). We attribute this decrease in ablation threshold to the increased concentration of free carriers in more highly doped silicon, increasing the availability of electrons able to take part in multiphoton and avalanche ionisation [11]. The narrowing of the bandgap is not thought to play a significant role here, as the bandgap only narrows significantly at dopant concentrations of around 10$^{18}$ cm$^{-3}$ or greater [16-19], whereas here we see reduction in the single pulse ablation threshold at dopant concentrations orders of magnitude lower than this. This reduction in ablation threshold is also seen by previous experiments by Cao and Fang et al. [24, 25], who also attribute the phenomenon seen to the initial concentration of free electrons.

There is also a marked difference seen between the single pulse ablation thresholds for N and P-type doped silicon (Figure 6). N-type doped silicon has a consistently lower single pulse ablation threshold than P-type, and the reduction in the ablation threshold with increasing dopant concentration is significantly greater for N-type. In fact, the reduction in ablation threshold between undoped and 0.001-0.005 Ω.cm doped silicon for N-type is double that seen for P-type. This is an interesting finding, and can be explained by the relative mobility of the different types of charge carriers involved. N-type silicon has free electrons available to carry charge, which have a much greater mobility in the lattice than do the electron holes that carry charge in P-type silicon [20, 35]. This allows the electrons to travel more freely within the lattice, increasing their utilisation for collisional (avalanche) ionisation. We hypothesise that this is the origin of the more efficient single pulse femtosecond laser micromachining for N-type Si over P-type.[36]

The infinite pulse ablation threshold ($F_{th,\infty}$ - Figure 6B) also gives some insight into the incubation behaviour and ablation characteristics of doped silicon. In contrast to the single pulse ablation threshold, we see no significant difference in the infinite pulse ablation threshold across the range of dopant types and concentrations studied. We attribute this observation to the accumulation of laser induced defects during machining. Femtosecond laser ablation is known to introduce defects at a level of around 10$^{22}$ - 10$^{23}$ cm$^{-3}$ [11], which is several orders of magnitude greater than the maximum defect concentration used in this study. As the laser-induced defects are far more numerous than the initial defects arising from doping, the main contributing factor towards the infinite pulse ablation threshold is the laser-induced defects rather than the dopant concentration (i.e. initial defect concentration).

These results are in contrast to the findings of Cao and Fang et al. [24, 25], who saw a difference in ablation threshold for different doping levels across all pulse numbers studied. Their study however, only reached a maximum of 1000 pulses, compared to the ~30,000 pulses maximum studied here, which may not have been sufficient to reach the constant value of the infinite pulse ablation threshold ($F_{th,\infty}$). We hypothesise that this steady state is reached due to the high concentration of laser-induced defects, leading to mechanical stresses within the solid which increase the energy barrier for formation of new defects [37]. The infinite pulse ablation threshold ($F_{th,\infty}$) thus corresponds to the pulse fluence that is no longer able to overcome this energy barrier and increase the defect concentration.

3.3 Incubation Parameters

In addition to examining the initial and final ablation thresholds, we also studied the incubation parameters obtained by fitting equation 3 to our data. Here we discuss not only the change in ablation threshold with increasing pulse number, but also how efficiently this change occurs.

The first parameter, $\Delta\alpha/\alpha_0$, describes the relative change in absorption from the initial ($\alpha_0$) to the final absorption coefficient ($\alpha$) [12]. The parameters $\beta$ and $\gamma$ describe the rate of the pulse-induced change (in absorption and critical fluence, respectively), using a similar mathematical form (exponential decay) to that suggested by Ashkenasi et al. [10]. All three of these parameters govern the change from the single pulse ablation threshold $F_{th,1}$ to the infinite pulse ablation threshold $F_{th,\infty}$.

In our experiments all values of $\Delta\alpha/\alpha_0$ are positive (Figure 7A), indicating that in all cases the laser ablation process results in increased absorption of the incident 800 nm light. This is unsurprising, and can be explained by a number of factors including increased roughness and defect concentration [38-40]. A clear trend can also be seen with respect to the dopant level in Figure 7A, with more highly doped silicon displaying a smaller change in absorption. This can be explained by the already high defect concentration in the doped silicon along with correspondingly higher absorption [39, 40], making further increases in defect concentration (and therefore absorption) more difficult. This trend for the change in absorption explains the different incubation behaviour seen in Figure 5, where the decrease in ablation threshold occurs later for more highly doped silicon. No significant differences in the change in absorption are seen between N-type and P-type doped silicon which is as expected, as the major factor for this absorption change in the defect concentration rather than the carrier mobility.

For $\beta$ and $\gamma$, the parameters controlling the efficiency of the change in absorption and critical fluence respectively, no significant difference is seen across the range of dopant concentrations and types studied (Figure 7B and C). This is to be expected, as pulse induced changes are mainly influenced by laser characteristics, with the defect concentration impacting the magnitude of these changes but not their rate. This may not be the case at dopant concentrations even higher than those studied here, where the laser's ability to introduce defects will be limited by the defects already in the structure; however the dopant concentrations studied here are low enough to avoid these effects.

4. Conclusions

Ablation threshold tests found that the increased free carrier concentration provided by doped silicon led to a decrease in the single pulse ablation threshold ($F_{th,1}$) with increasing dopant concentration; this decrease was greater for N-type doping than P-type due to the higher mobility of electrons compared to electron holes allowing greater utilisation of charge carriers for collisional ionisation. This led to a maximum decrease in the single pulse ablation threshold ($F_{th,1}$) of almost 30% (compared to undoped silicon) for silicon highly doped with phosphorus.

No significant change was observed in the infinite pulse ablation threshold ($F_{th,\infty}$) across all the doping levels and types, owing to the large amount of defects induced by the laser ablation process, which is orders of magnitude greater than the dopant concentration, making laser induced defects by far the dominant factor affecting the infinite pulse ablation threshold.

Some differences in incubation behaviour are seen across different doping levels, with more highly doped silicon requiring more pulses before reduction in the ablation threshold occurs. This is accounted for by the reduced amount of absorption change possible in more highly doped silicon.

Ultrafast laser ablation thresholds and incubation effects for doped silicon were measured across a large range of dopant types, concentrations, and pulse numbers. These results show the effect of dopant concentration on ablation threshold and incubation, and we propose a mechanism by which these effects occur. This provides a sound basis for future experiments of laser ablation of doped silicon. In addition, the diagonal scan (D-Scan) ablation threshold measurement technique was utilised, further establishing this valuable technique's effectiveness.

5. Acknowledgements

The authors acknowledge financial support from the New Zealand Ministry of Business, Innovation and Employment Grants UOAX1202 (Laser Microfabrication and Micromachining) and UOAX1416 (Tailored Beam Shapes for Fast, Efficient and Precise Femtosecond Laser Micromachining).

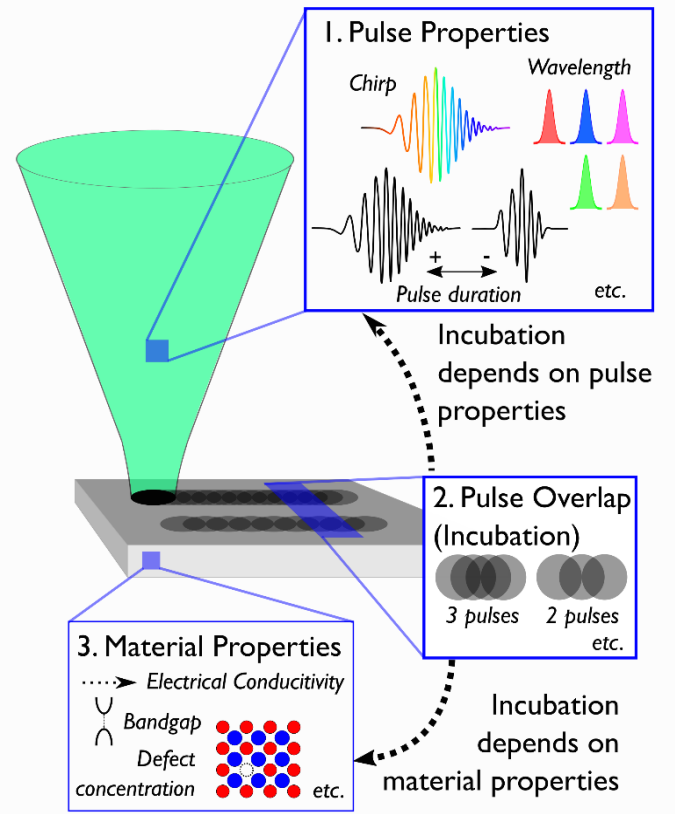

*Figure 1: Schematic diagram of the factors affecting the ultrafast laser ablation threshold, indicating that material properties, pulse properties, and number of pulses all individually affect the ablation threshold. In addition, the way in which the number of pulses affects the ablation threshold (the phenomenon known as incubation) is in turn affected by the material and pulse properties.*

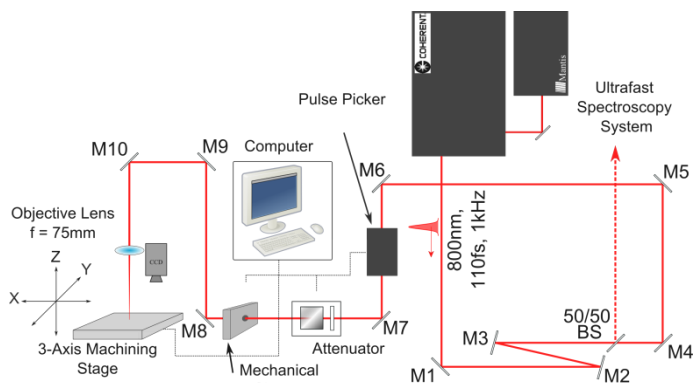

*Figure 2: Diagram of the laser setup used for these experiments. The laser source is an oscillator and amplifier (Mantis and Legend Elite, both Coherent Inc., USA) operating at 800 nm, 110 fs, 1 kHz. The beam is directed using ultrafast mirrors and a 50:50 beam splitter, to a pulse picker (Series 5046ER, Fastpulse Inc., USA) and attenuator (Watt Pilot, UAB Altechna, Lithuania), then through a 75 mm focal length lens (LA1608-B-ML, Thorlabs Inc., USA) onto the 3-axis micromachining stage (Thorlabs Inc., USA).*

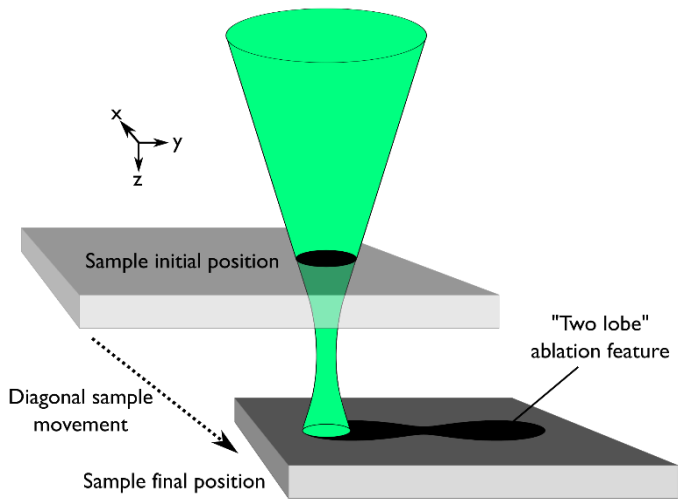

*Figure 3: Illustration of the diagonal scan ablation threshold measurement method, showing the sample being translated diagonally through the focal point.*

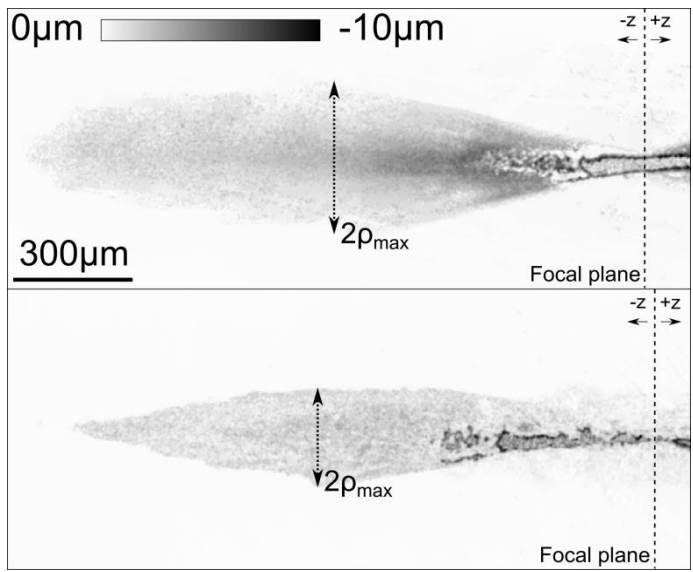

*Figure 4: Example surface height plots obtained using the optical profiler, showing examples of diagonal scan laser ablation features machined into silicon. Top: Undoped silicon, bottom: P-type doped silicon, 0.01-0.05 Ω.cm. We ignore features made below the focal plane (+z region) as in this area the beam is distorted by plasma generated at the focal point.*

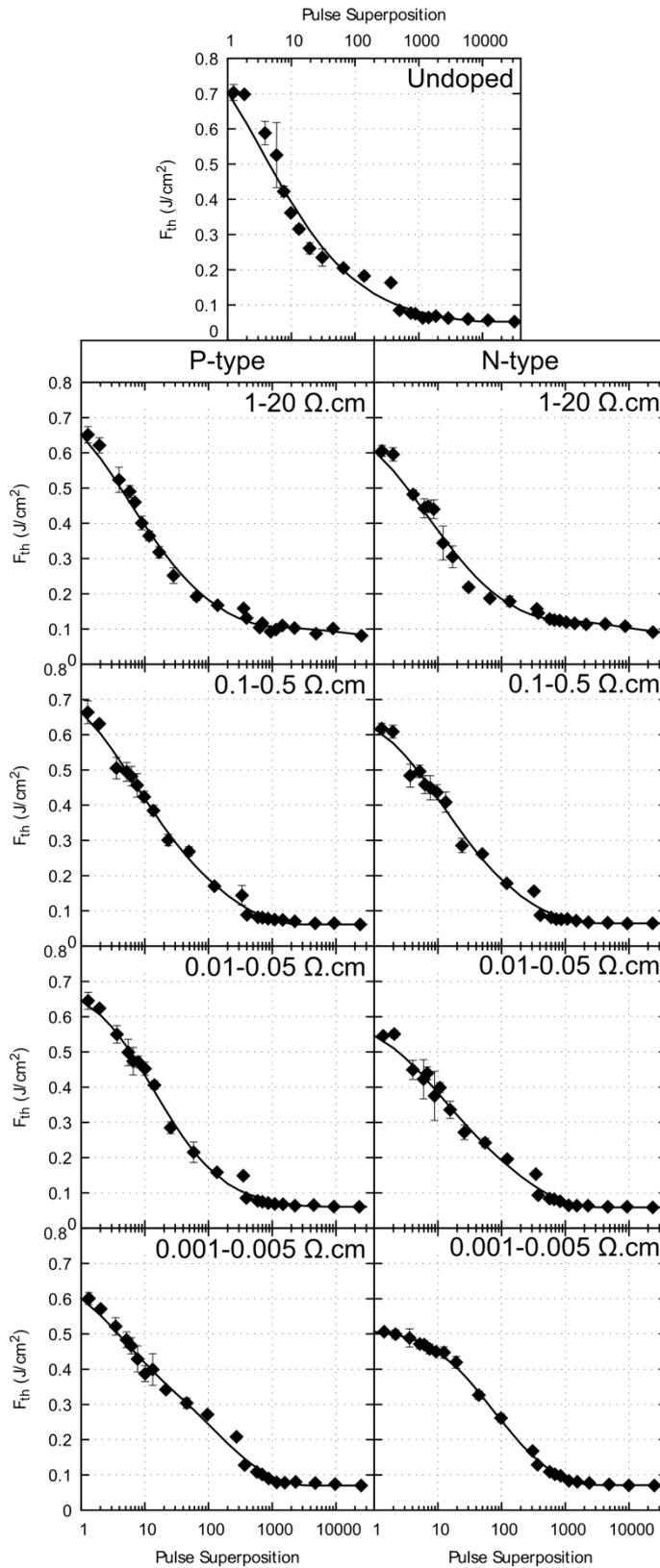

*Figure 5: Incubation curves (ablation threshold vs. pulse superposition) for doped and undoped silicon, fitted to equation 3. See Table 1 for details of resistivity and dopant concentration values. Ablation tests carried out using the D-Scan method, with 110 fs, 800nm pulses, at repetition rates from 4 – 1000 Hz.*

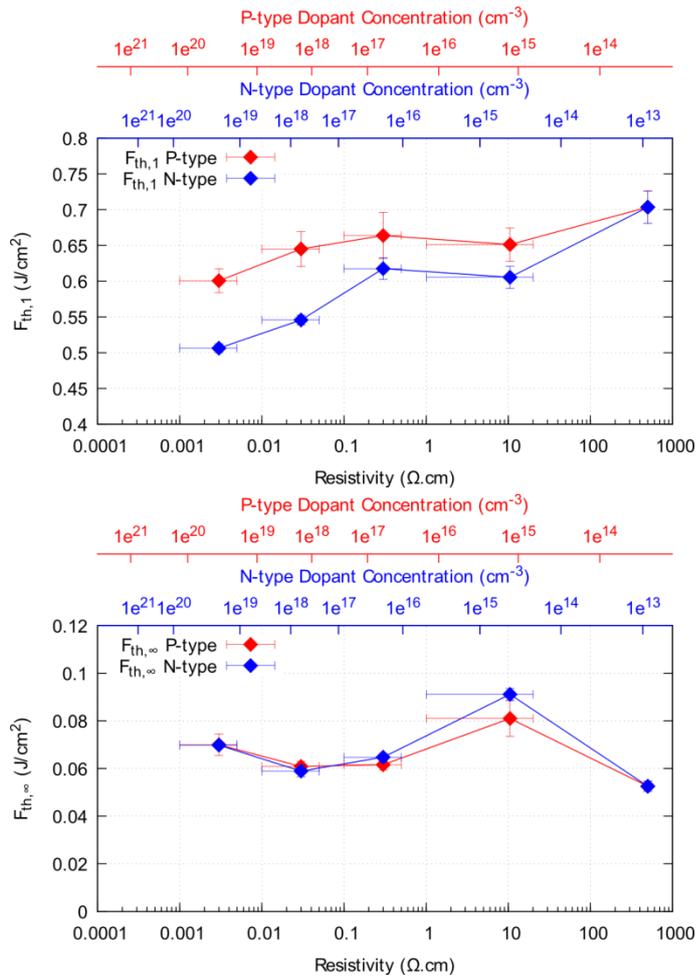

*Figure 6: Ablation thresholds across the range of doped silicon samples measured, showing top: single pulse ablation threshold and bottom: infinite pulse ablation threshold.*

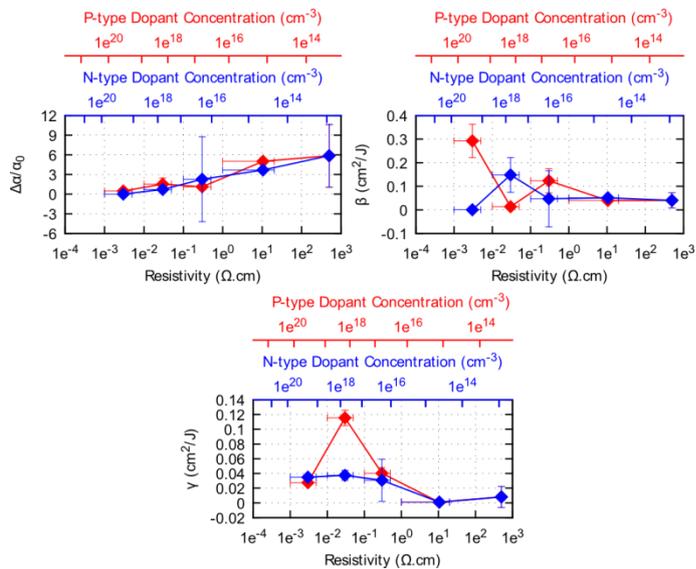

*Figure 7: Incubation curve fit parameters, top left: change in absorption $\Delta\alpha/\alpha_0$, top right: efficiency of absorption change $\beta$, and bottom: efficiency of critical fluence change $\gamma$. Parameters shown across the range of doping levels tested, for N-type (blue) and P-type (red) doping.*